\documentclass[prl,twocolumn,amsmath,amssymb,floatfix,superscriptaddress]{revtex4-1}

\usepackage{graphicx}
\usepackage{amssymb}
\usepackage{amsmath}
\usepackage{color}
\usepackage{ wasysym }
\usepackage{hyperref}

\def\barray{\begin{array}}
\def\earray{\end{array}}
\def\be{\begin{equation}}
\def\ee{\end{equation}}
\def\ben{\begin{equation} \nonumber}
\def\een{\end{equation}}
\def\ban{\begin{eqnarray*}}
\def\ean{\end{eqnarray*}}
\def\ba{\begin{eqnarray}}
\def\ea{\end{eqnarray}}

\def\({\left(}
\def\){\right)}
\def\half{{1\over2}}

\def\axion{\mathcal{X}}

\begin{document}

\title{Chromo-Natural Inflation:\\ Natural inflation on a steep potential with classical non-Abelian gauge fields}
\author{Peter Adshead}
\affiliation{Kavli Institute for Cosmological Physics,  Enrico Fermi Institute, University of Chicago, Chicago, Illinois 60637, U.S.A}
\author{Mark Wyman}
\affiliation{Kavli Institute for Cosmological Physics,  Enrico Fermi Institute, University of Chicago, Chicago, Illinois 60637, U.S.A}
\affiliation{Department of Astronomy \& Astrophysics, University of Chicago, Chicago, Illinois 60637, U.S.A.}

\begin{abstract}
We propose a model for inflation consisting of an axionic scalar field coupled to a set of three non-Abelian
gauge fields. Our model's novel requirement is that the gauge fields begin inflation with a
rotationally invariant vacuum expectation value (VEV) that is preserved through identification of SU(2) gauge invariance
with rotations in three dimensions.  The gauge VEV interacts with the background value of the axion, leading
to an attractor solution that exhibits slow roll inflation even when the axion decay constant has a natural value
($<M_{\rm Pl}$). Assuming a sinusoidal potential for the axion, we find that inflation continues until the axionic potential
vanishes. The speed at which the axion moves along its potential is modulated by its interactions with the gauge VEV, rather than being determined by the slope of its bare potential. For sub-Plankian axion decay constants vanishingly small tensor to scalar ratios are predicted, a direct consequence of the Lyth bound.  The parameter that controls the 
interaction strength between the axion and the gauge fields requires a technically natural tuning of $\mathcal{O}$(100). 
\end{abstract}
\maketitle

The success of the inflationary paradigm is manifest. Observations, preeminently of the cosmic
microwave background radiation, have confirmed that our Universe is approximately spatially flat and have provided compelling evidence that structure formation was initiated with nearly scale invariant curvature fluctuations over the entire range of observable 
scales. The data suggest that these fluctuations are Gaussian and are red-tilted, in perfect concord with the simplest models of inflation. 

Despite these successes, many physicists believe that the simplest models of inflation are, on their own,
inadequate. This is because many successful models rely on scalar fields whose potentials are tuned to
extraordinary flatness to achieve sufficient inflation. This tuning is not generically protected from quantum
corrections. Many models of inflation now exist that utilize a variety of methods to alleviate this difficulty. 
Among these methods, a particularly natural possibility is protection by a shift symmetry, as we find in axions.
Natural Inflation was the first model to make use of an axionic shift symmetry in this context \cite{Freese:1990rb}.  While Natural inflation is observationally viable, matching observations requires a large axion decay constant, $f \sim M_{\rm Pl}$ \cite{Freese:2004un}.
Such a large decay constant appears to be difficult to realize in string theory \cite{Banks:2003sx}.
The discovery of viable axionic inflationary models that do not need super-Planckian decay constants (e.g., \cite{Silverstein:2008sg, Kim:2004rp, Barnaby:2010vf}) has led to a
renaissance of interest in axionic inflation.

In this letter, we propose a new method for inflating with an axion that has a sub-Planckian decay constant. Our model's new ingredient is a collection of non-Abelian gauge fields with a vacuum expectation value (VEV).  Since every SU(N) group has an SU(2) subgroup, SU(N) gauge fields can always be given a VEV that is rotationally invariant by identifying the global part of the SU(2) symmetry  with the rotational symmetry of space (for more details, see \cite{Galtsov:1991un,Gal'tsov:2010dd} and references therein)\footnote{The authors of \cite{Maleknejad:2011jw,Maleknejad:2011sq} found that inflation can proceed if the gauge VEV equation of motion is dominated by a self-interaction term of the form $ (F\tilde{F})^2$; similar configurations are studied in \cite{Gal'tsov:2010dd,Gal'tsov:2011dm}; see also \cite{Alexander:2011hz} . These scenarios require higher dimension operators to dominate dimension four operators. We can realize such a situation in our model by integrating the axion near the bottom of its potential. In this sense, the model of \cite{Maleknejad:2011jw, Maleknejad:2011sq} is equivalent to ours near the bottom of the axion potential.}. While our proposal does not rely on the specific gauge group, we work with SU(2) gauge fields for concreteness.

This model produces successful inflation in a new way, achieving slow roll inflation by the efficient  transfer of 
axionic energy into classical gauge fields, rather than from dissipation via Hubble friction. This energy exchange mediated by the coupling between the axion and the Chern-Simons term for the non-Abelian gauge field. Because the Chern-Simons term, by itself, is a total derivative, this coupling respects the axionic shift symmetry. Hence, any other axion-gauge field interactions will be absent, while higher-order
corrections to the gauge action will be very small. This means that we can consistently tune the coefficient of one operator (the axion-Chern-Simons term) to be large in a technically natural way. This is in contrast to generic single field models, where the  form of the Lagrangian is not protected by any symmetry. In those models, a tower of fine tunings are needed  to enforce cancellations of one-loop effects.


{\bf Inflationary dynamics:} To demonstrate our model's viability, we look for inflationary solutions of the axion-gauge system in the regime where the axion decay constant is small, $f \ll M_{\rm pl}$. We consider the action for an axion interacting with a set of SU(2) gauge fields, 
\begin{align}\label{eqn:action}
\mathcal{L} = \sqrt{-g} & \left[-\frac{R}{2}-\frac{1}{4}F_{\mu\nu}^{a}F_{a}^{\mu\nu}  - \frac{1}{2}(\partial \axion )^2\right . \nonumber \\
& \left . - \mu^4\left(1+\cos\left(\frac{\axion }{f}\right)\right)-\frac{\lambda}{8f} \axion  F^{a}_{\mu\nu}\tilde F_{a}^{\mu\nu}\right],
\end{align}
where we work in natural units ($\hbar = c = M_{\rm pl} = 1$), Greek letters denote spacetime indices while Roman letters from the start of the alphabet denote gauge indices, and Roman letters from the middle of the alphabet denote spatial indices. In this expression, the field $\axion$ is an axion which we take to be in a homogeneous configuration,
$
\axion = \axion(t)$. Note that although we have written the standard cosine potential for the axion -- which arises due to non-perturbative effects from the interaction of the axion with some other gauge sector -- nothing about our mechanism relies on the potential taking this particular form.
As we mentioned before, we assume an initial gauge field configuration described by a rotationally invariant, classical VEV \footnote{We have verified that this field configuration is classically stable against homogeneous fluctuations  \cite{PAMWnext}. It is also known that classical configurations of chromo-electric and magnetic fields are unstable to quantum fluctuations within the context of QCD \cite{Nielsen:1978rm,Savvidy:1977as}.  However, our case is different enough that these QCD instabilities are not immediately applicable to our case.},
\begin{align}
A^{a}_{0} = 0,\quad A^{a}_i  =\psi(t) \,a(t)\delta^{a}_i.
\end{align}
The related field strength tensor has components,
\begin{align}
F_{0i} = \partial_t (\psi(t) \,a(t))\delta^{a}_{i}, \quad \quad
F^{a}_{ij} =  -\tilde g f^{a}_{ij}(\psi(t) \, a(t))^2,
\end{align}
where the $f^{a}_{ij}$ are the structure functions of SU(2), an overdot denotes a derivative with respect to cosmic time, and $\tilde g$ is the gauge coupling. 
With this field configuration, the reduced action for these degrees of freedom is,
\begin{align}
\mathcal{L}_m = \nonumber & a^3 \left [  \frac{3}{2} \frac{1}{a^2} \(\frac{\partial (\psi a)}{\partial t}\)^2 - \frac{3}{2}\tilde g^2 \psi^4+ \frac{1}{2}\dot \axion^2 \right . \\  & \left . - \mu^4\left(1+\cos\left(\frac{\axion }{f}\right)\right)   -3 \tilde g \frac{\lambda}{f} \axion \frac{\psi^2}{a} \frac{\partial ( \psi a)}{\partial t} \right],
\end{align}
from which the equation of motion for the axion is,
\begin{align}\label{eqn:axioneom}
\ddot \axion + 3H \dot\axion -\frac{\mu^{4}}{f}\sin\left(\frac{\axion}{f}\right)= &  -3 \tilde g \frac{\lambda}{f}\psi^2\(\dot\psi + H\psi\),
\end{align}
and the equation of motion for the gauge field is
\begin{align}
\label{eqn:psieom}
\ddot{\psi} + 3H\dot\psi+ (\dot H+ 2H^2) \psi  + &  2\tilde g^2 \psi^3 =  \tilde g  \frac{\lambda}{f}  \psi^2 \dot\axion.
\end{align}
The gauge field has the equation of state of radiation, $p = \rho/3$, and thus by itself cannot source inflationary expansion, while the gauge field-axion interaction term does not contribute to either the energy density or pressure. 
The system thus inflates if the energy density is dominated by axion potential. In the absence of the gauge fields, the model reduces to Natural Inflation; in the absence of the coupling to the axion, the gauge VEV decays within a Hubble time. The key new ingredient in our scenario is that, when the gauge field is turned on in the homogeneous configuration described above, the equation of motion for the axion -- Eqn.\ (\ref{eqn:axioneom}) --  gains a new 
interaction term on the right hand side. As we will see, this provides an additional lever with which we can slow
the axion's evolution, making it possible for slow roll inflation to occur away from the hilltop even for sub-Planckian axion decay constants, $f < M_{\rm pl}$ \footnote{Our proposal is similar to that of  \cite{Anber:2009ua}, who achieve a phase of slow roll inflation on a steep axion potential via electro-magnetic dissipation. 
While  \cite{Anber:2009ua} relies on a quantum back-reaction of the photons on the axion condensate to slow the rolling, our proposal imparts this effect at the classical level. Nonetheless, it is important to check whether the particle production that leads to difficulties in their model is under control in ours. In the model at hand, the key parameter, $\xi$, is below the bound of $\xi<4.7$ found in \cite{Barnaby:2011qe} (here $\xi \simeq 2.5$) during the observable epoch. While the model at hand is non-Abelian, since we work in the weakly coupled regime, we expect similar bounds to apply. We note that this is a value associated with observable non-Gaussianities in the treatment of \cite{Barnaby:2011qe}}. 

We can now look for slowly rolling inflationary solutions of this system of equations assuming $\ddot \axion$, $\ddot \psi$, and $\dot H \simeq 0$
and studying the resulting equations. In this limit, we can diagonalize the system, finding equations for $\dot \psi$ and $\dot \axion$ that depend
only on $\psi$ and $\axion$. The equation for $\psi$ that results from this procedure reads
\begin{align}\nonumber
\dot \psi = & - H\frac{ \psi( 2 f^2 H^2  +  2 \tilde g^2 f^2 \psi^2 + \tilde g^2 \lambda^2 \psi^4) }{ \( 3f^2 H^2+ \tilde g^2 \lambda^2 \psi^4\)}\\ &\qquad +\frac{ \tilde g \lambda \mu^4 \psi^2 \sin(\axion/f)}{3 \( 3f^2 H^2+ \tilde g^2 \lambda^2 \psi^4\)}.
\end{align}
Choosing parameters that make
$
3f^2 H^2 \ll \tilde g^2 \lambda^2 \psi^4
$
gives
\begin{align}\label{eqn:psidot}
H\dot \psi & \simeq -H^2 \psi + \frac{\mu^4 \sin(\axion/f)}{3 \tilde g\lambda}\frac{H}{\psi^2}.
\end{align}
This parameter choice amounts to making the combination $\lambda/f$ to be large, since the inflationary solution requires $\tilde g^2 \psi^4 \ll \mu^4 \sim H^2$; see appendix B of \cite{Anber:2009ua} for several examples of how to achieve a large value of $\lambda$. Assuming we are in  the overdamped regime, the right hand side of Eqn.\ (\ref{eqn:psidot}) can be interpreted as the slope of an effective potential for the gauge field,
\be
V_{\mbox{eff}}(\psi) =H^2 \frac{\psi ^2}{2}+ \frac{\mu ^4 \sin \left( \axion/f \right)}{3 \tilde g  \lambda }\frac{H}{\psi}.
\ee 
This effective potential has a minimum:
\begin{align}\label{eqn:slowrollpsi}
\psi_{\rm min} \approx \(\frac{\mu^4 \sin(\axion/f)}{3 \tilde g\lambda H}\)^{1/3}.
\end{align}
When the gauge VEV takes this value (and $\dot \psi \simeq 0$) the right hand side of Eq. \ref{eqn:axioneom} is exactly equal to the gradient of the bare axion potential. What this means is that the axion's motion is dominated
by classical energy transfer into the gauge sector. The axion still rolls, but only very slowly;
since we have diagonalized our system of equations, the appropriate equation to consider is not Eq. \ref{eqn:axioneom}, but the diagonalized one, which we write explicitly below (Eq. \ref{axioneqn}).
The mass of fluctuations about the gauge VEV's minimum are large ($m^2_{\psi}= 3 H^2$). Thus, when the gauge VEV has settled into this minimum we can set $\dot \psi \simeq 0$ and integrate out $\psi$  during slow roll. This gives us an effective equation governing the axionÕs evolution in terms
of X alone. We emphasize that it is important to diagonalize the equations before making the replacement $\psi = \psi_{\rm min}$.
Working with efolding number, $dN = H dt$, rather than time, we find
\be
\label{axioneqn}
\frac{\axion'}{f} \simeq \frac{2}{\sqrt[3]{3^2}} \( \frac{3 \lambda^{2/3} H^{8/3} + \sqrt[3]{3} \tilde g^{4/3} \mu^{8/3} \sin^{2/3} \(\axion/f \)}{\sqrt[3]{\tilde g^2 \lambda^4 \mu^4 H^4 \sin \( \axion/f \) }} \),
\ee
where a prime ($'$) denotes a derivative with respect to $N$. We can see from this equation that the axion's dynamics are not being dictated by the interplay of the gradient of the axion's bare potential and Hubble damping;
instead, the axion is evolving quite slowly on a flat effective potential generated by interaction of the axion and the gauge VEV. 
We note also in passing that $\axion' \propto 1/\sin^{1/3}(\axion/f)$, which diverges mildly as $\axion \rightarrow 0$. This is indicative of the fact that the gauge field provides a force on the axion even when $\axion =0$, implying that inflation cannot continue indefinitely at the hilltop even classically when the gauge VEV is present. 
This formula can be inverted to find an expression for the number of efoldings over which inflation proceeds. Assuming inflation begins at $\axion_0$, inflation will end at $\axion/f \equiv x = \pi$.  Taking $H$ to be dominated by the axion potential and writing $\tilde \mu \equiv \mu /M_{\rm pl}$, the formula for the number of e-foldings, $N$, is given by
\be\label{eqn:efoldings}
N(\axion_0) = \int_{\frac{\axion_0}{f}}^\pi \hspace{-4 pt} \frac{\half \(3 \tilde  g^{2} \lambda ^{4} \tilde \mu^{4} \left( 1+\cos  x \right)^{2} \sin x\)^{1/3}}{ \( \lambda ^{2} \tilde \mu^{8} \left( 1+\cos x \right)^{4}\)^{1/3}+  \(3 \tilde g^{2} \sin x\)^{2/3} } dx.
\ee
The combination of parameters that maximizes the number of efolds of inflation $N$ is given by 
$\tilde g^2 /\lambda \simeq \tilde \mu^4/3$.
Assuming this relation,  a simple expression can be found for the maximum number of efoldings
\be
N_{\rm max} \simeq (3/5) \lambda \label{efolds},
\ee
where the numerical prefactor comes from integrating the combination of sine and cosine functions over $x \in [0,\pi]$.
Hence we can see that $\lambda\sim100$ is necessary to get $\mathcal{O}(60)$ efoldings of inflation.

The kinetic energy of the axion is never important in this model and the kinetic energy of the gauge field is negligible, and so the slow roll parameters are given by
\begin{align}
\epsilon &  \equiv -\frac{H'}{H} 
\simeq   \frac{3 \tilde g^2 \psi^4}{\mu^{4} (1+\cos(\axion/f))}+\psi^2, \nonumber
\end{align}
where we have used that $H$ is dominated by the axion potential; and
\be\label{eqn:eta}
\eta \equiv \frac{\epsilon'}{\epsilon} \simeq 2\tilde g^2 \frac{\psi^4}{H^2} + \frac{\psi'}{\epsilon}\(12 \tilde g^2\frac{\psi^3}{H^2}+2 \psi\).
\ee
Equation (\ref{eqn:eta}) implies that, during successful slow roll, the second term 
must be $\mathcal{O}(\epsilon^2)$. This happens when we pick parameters that can achieve enough inflation.
Although the slow roll parameters do not come into our calculations directly, it's worth noting that we have $\epsilon \propto 1/\lambda$
when the maximization condition is met. When the gauge VEV is at  its minimum ($\psi'=0$), $\eta = \mathcal{O}(\epsilon)$.

{\bf Estimating Perturbations:} The classical evolution of the system undergoing slow roll is, to an excellent approximation, described by the axion rolling slowly along a dynamically generated effective potential and the gauge VEV $\psi$ sitting at its dynamically enforced minimum, Eqn. (\ref{eqn:slowrollpsi}). Moreover, fluctuations of the gauge field about the minimum have mass $m_{\psi}\sim \sqrt{3} H$, from which it appears that they may play only a limited role in the adiabatic perturbations of the model. With this in mind, we can get a rough estimate for the amplitude and tilt of the power spectrum of adiabatic perturbations by assuming that the curvature perturbation is dominated by quantum fluctuations of the axion along the classical trajectory generated by its interactions with the gauge VEV. This path can be thought of as a kind of dynamically generated effective potential. The amplitude of the perturbations can then be estimated by considering the perturbation in the efolding number
\begin{align}\label{eqn:curvature}
\delta N   =  \frac{\partial N(\axion_0)}{\partial{\axion_0}}\delta\axion+\ldots,
\end{align}
where we have neglected the contributions of fluctuations of the gauge fields as discussed above. 
That is, for a simple estimate we can consider the axion's evolution along its effective potential to set the clock for the time evolution of the system. 
Hence, to a first approximation the power spectrum of curvature fluctuations will be given by
\begin{align}\label{eqn:power}
\Delta^{2}_{\mathcal{R}}(k) \sim \frac{1}{4\pi^2}\frac{H^2}{\axion'^2}.
\end{align}
Under the same set of assumptions, the tilt of the scalar spectrum, $n_s$, is given by
\be
n_s  \sim 1 - 2 \epsilon + 2 \frac{\axion''}{\axion'}.
\ee
One can verify that the COBE normalization can be matched along with a red spectral index with $n_{s} \approx 0.97$ at 50 efolds before inflation ends with the parameter choices $\{\mu,f,\tilde g, \lambda\} = \{3.16 \times 10^{-4}, 0.01, 2.0 \times10^{-6},  200.\}$. Let us stress that these are merely naive estimates of the fluctuations. Nonetheless, they indicate that the model is potentially viable.

We expect that the spectrum of tensor fluctuations in this model will be dominated by the usual vacuum fluctuations of the metric, so we can assume that its amplitude is set by the inflationary energy scale. While spin-2 excitations of the gauge field configuration may also contribute to the gravitational wave spectrum, we anticipate that they will be sub-dominant \cite{Maleknejad:2011jw, Maleknejad:2011sq}. This then implies that, in the absence of non-adiabatic evolution of the system, super-Planckian axion decay constants are a \emph{necessary} condition in order to yield observable gravitational waves. This is a simple consequence of the Lyth bound \cite{Lyth:1996im}. In this model, inflation ends very near a particular point on the potential, namely $\axion/f \approx \pi$. As the axion decay constant $f$ is decreased, the distance the field rolls in field space is correspondingly smaller by the same fraction. This means the field must roll more slowly in order that sufficient inflation is generated. The position of the field on the effective potential measures the time before inflation ends in this model, as is usual in models of inflation with a single effective adiabatic degree of freedom, and the curvature perturbation is the fluctuation of this time due to the quantum fluctuations in the metric and fields. Since the field is rolling more slowly, the size of the quantum fluctuations must also be smaller in order that the ratio in Eqn.\ (\ref{eqn:power}) satisfy the COBE normalization. For the parameters chosen above, $r \sim 10^{-6}$, which is far below the range probed by current or planned future experiments.

{\bf Discussion and Conclusions:}
We have described a new model for inflation driven by an axion. In our model, the novel ingredient is a collection of non-Abelian gauge fields that have an isotropic vacuum expectation value as their initial state. We identify the global SU(2) gauge symmetry with rotations in space, thus protecting the isotropy of the gauge VEV with its SU(2) gauge symmetry. The
gauge VEV, in turn, replaces Hubble friction as the mechanism by which the axion's speed of rolling down its potential is slowed. The equations that
govern the gauge VEV have a dynamical minimum. When the VEV settles into this minimum, the resulting effective action for the axion
is dominated by classical gauge field interactions. The slow evolution of the axion along this classical trajectory is similar
to classical evolution along a flat potential, suggesting that the
adiabatic perturbations of this model will be compatible with observations. Meanwhile, fluctuations in
the gauge VEV, orthogonal to the adiabatic trajectory, have effective masses of order the Hubble scale. This suggests that we can estimate the amplitude and tilt of the scalar fluctuations by assuming that quantum fluctuations of the position of the axion on its effective potential (the clock in our model) dominate the curvature perturbation. Fluctuations off this trajectory will decay outside the horizon, and are thus expected to contribute negligibly. Obtaining sufficient inflation while matching the observed amplitude and tilt of the perturbations provides three constraints on combinations of the four parameters of this model leaving one degree of freedom.

The effective theory we study requires a single, technically natural tuning: the coupling between the axion and the gauge fields via the Chern-Simons term must be large. The shift symmetry of the axion and the restrictions of non-Abeliean
gauge interactions guarantee that we have neglected no important corrections to our effective action.
It is also easy to see that the energy scale of inflation is well below the cut-off of the effective theory, 
since the cut-off is given by $f / \lambda \gg H \simeq \mu^2$.   
 Isocurvature and statistics beyond the power spectrum -- such as the bispectrum -- may provide unique signatures of the mechanism and further bound the allowed region of parameter space. We caution, however, that we have made mere estimations of the perturbations in this theory. In practice, a full analysis of the six extra degrees of freedom relative to the standard inflationary scenario is necessary to measure the health of this theory relative to current data. This is greatly complicated by the addition of the gauge fields, which mix with the axion at tree level and must satisfy a non-Abelian version of Gauss' law.  We will provide a complete account of the perturbations in a future publication \cite{PAMWnext}.

\acknowledgements

We thank Neil Barnaby,  Adam Brown, Cliff Burgess, Rich Holman, Wayne Hu, Emil Martinec, David Seery, and Sav Sethi for useful conversations. This work was supported in part by the Kavli Institute for Cosmological Physics at the University of Chicago through grants NSF PHY-0114422 and NSF PHY-0551142 and an endowment from the Kavli Foundation and its founder Fred Kavli.  MW was supported by U.S. Dept. of Energy contract DE-FG02-90ER-40560
\appendix

\bibliography{agf}

\end{document}